\documentclass[
    ,final            
    ,numberedheadings 
  ,bibliocites
  ]
  {aipproc}

\layoutstyle{6x9}

\usepackage{natbib}

\begin{document}
\title{Photoevaporation of Minihalos during Reionization}

\author{Paul R. Shapiro}{
address={Department of Astronomy, University of Texas at Austin}
}

\author{Ilian T. Iliev}{
address={Osservatorio Astrofisico di Arcetri, Italy}
}

\author{Alejandro C. Raga}{
address={Instituto de Ciencias Nucleares, UNAM, Mexico}
}

\author{Hugo Martel}{
address={Department of Astronomy, University of Texas at Austin}
}


\def\gtrsim{\mathrel{\raise.3ex\hbox{$>$}\mkern-14mu
             \lower0.6ex\hbox{$\sim$}}}
\def\lesssim{\mathrel{\raise.3ex\hbox{$<$}\mkern-14mu
             \lower0.6ex\hbox{$\sim$}}}

\begin{abstract}
We present the
first gas dynamical simulations of the 
photoevaporation of cosmological minihalos
overtaken by the ionization fronts which swept
through the IGM during reionization in
a $\Lambda$CDM universe, including the
effects of radiative transfer. We
demonstrate the phenomenon of I-front trapping
inside minihalos, in which the weak, R-type fronts
which traveled supersonically across the IGM 
decelerated when they encountered the dense,
neutral gas inside minihalos, becoming D-type I-fronts, preceded by
shock waves. For a minihalo with virial
temperature $T_{\rm vir}\leq10^4\rm K$, the I-front gradually
burned its way through the minihalo which trapped
it, removing all of its baryonic gas by causing a
supersonic, evaporative wind to blow backwards
into the IGM, away from the exposed layers of
minihalo gas just behind the advancing I-front.

Such hitherto neglected feedback effects were
widespread during reionization. N-body
simulations and analytical estimates of halo
formation suggest that sub-kpc
minihalos such as these, with
$T_{\rm vir}\leq10^4\rm K$, were so common as to dominate the
absorption of ionizing photons. This means that previous estimates of
the number of ionizing photons per H atom required
to complete reionization which neglected this
effect may be too low.  Regardless of their effect
on the progress of reionization, however, the
minihalos were so abundant that random lines of
sight thru the high-$z$ universe should encounter
many of them, which suggests that it may be
possible to observe the processes described here
in the absorption spectra of distant sources.  
\end{abstract}
\maketitle

\begin{figure}[b!] 
\vspace{-70pt}
  \includegraphics[width=4.9in]{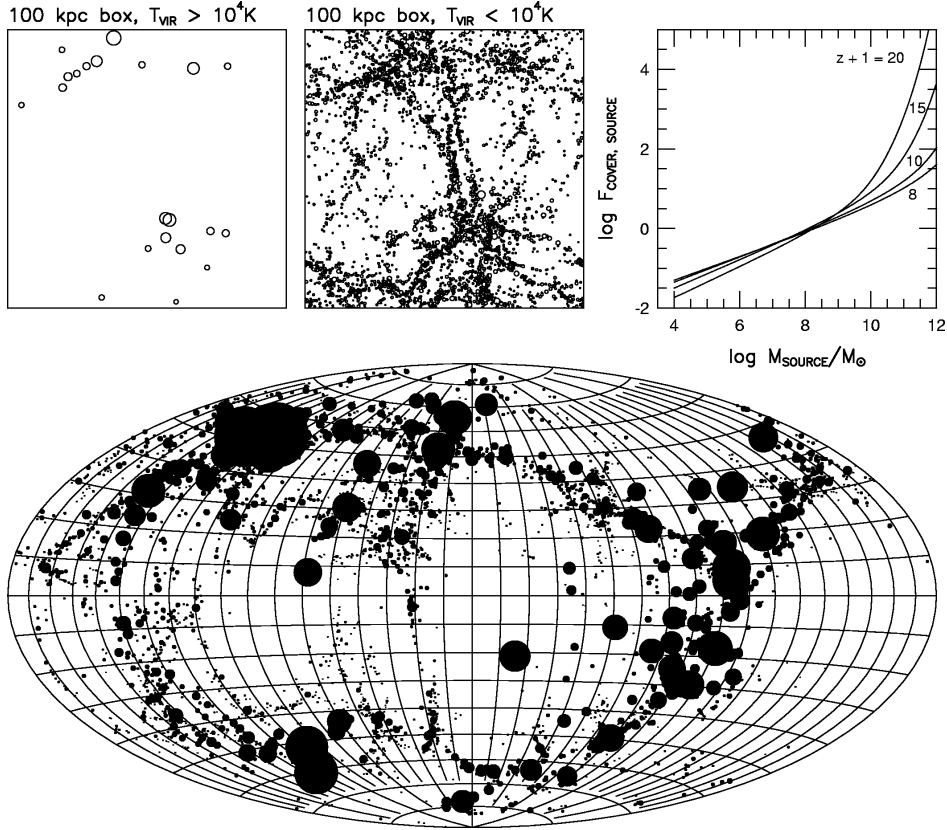}
\vspace{-180pt}
\caption{
Dark matter halos of mass $M>10^{7.6}M_\odot$ (i.e. $T_{\rm vir}>10^4K$)
(a) (top left) and $10^{7.6}M_\odot>M>10^{5.6}M_\odot$ 
(i.e. $T_{\rm vir}<10^4\rm K$) (b) (top center)
found in $\rm P^3M$ simulation of $\Lambda$CDM universe in 100 kpc (proper) box,
at $z=9$.
(c) (top right) Fraction of sky covered by minihalos located within the
mean volume per source halo versus halo mass,
at different redshifts, computed using the TIS model and the PS approximation,
corrected for linear bias. 
(d) (bottom) Sky covering by minihalos as seen from a 
$1.1\times10^8M_\odot$
source, from $\rm P^3M$ simulation with 50 kpc box at $z=9$.
All minihalos within $\rm 25\,kpc$  are plotted.
The covering fraction is 23.6\%.}
\label{fig1}
\end{figure}

\begin{figure}[b!] 
\vspace{-175pt}
  \includegraphics[width=5.2in]{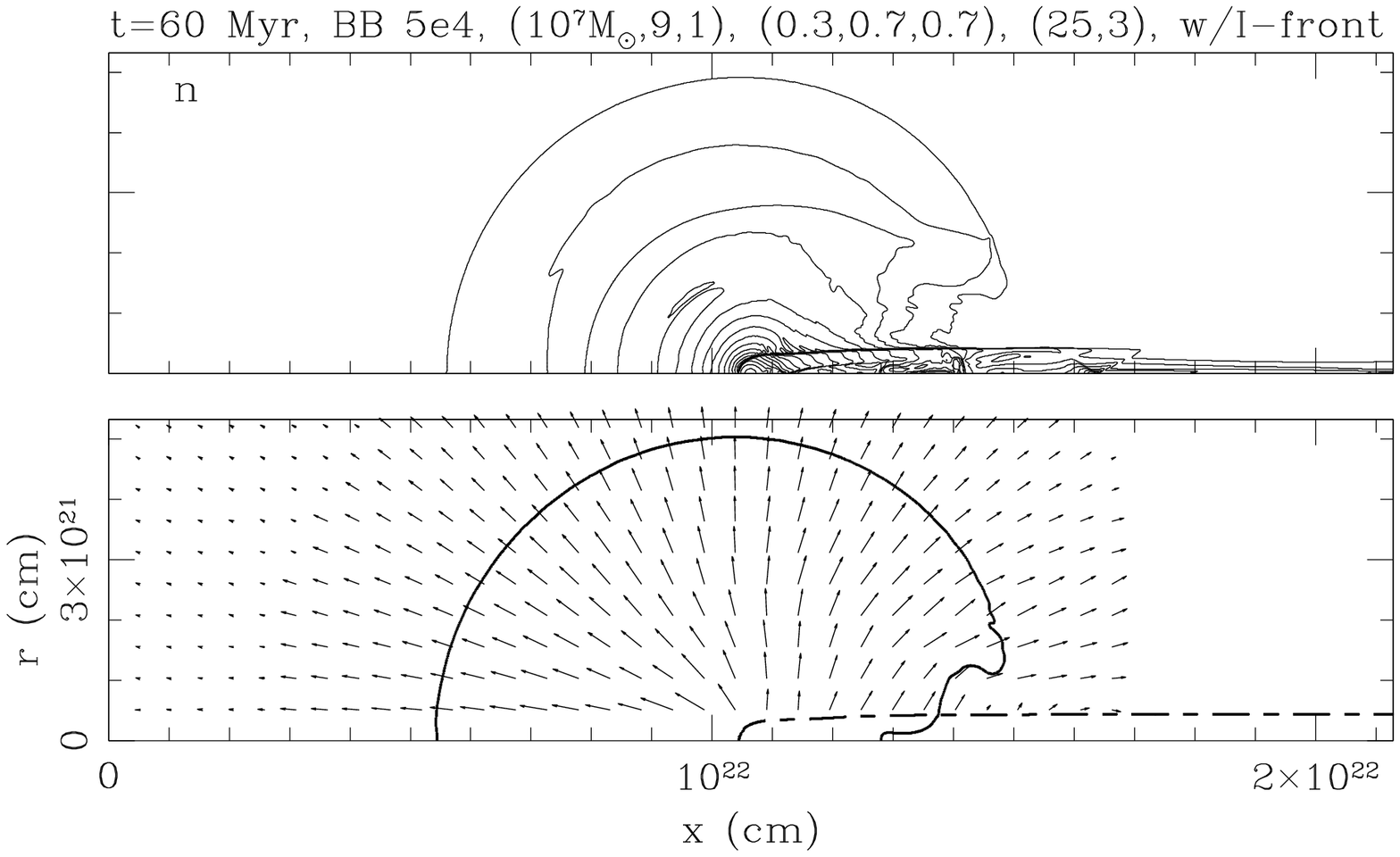}
\vspace{-125pt}
\caption{One time-slice, 60 Myr after I-front caused by source (located far
to the left, along the $x$-axis) overtakes a
$10^7M_\odot$ minihalo [centered at $(r,x)=(0,1.06\times10^{22}\,\rm cm$)]
at $z=9$ in the $\Lambda$CDM universe, for source with $F_0=1$, a
stellar BB spectrum $T_*=50,000\,\rm K$. (a) (top) 
Isocontours of atomic density,
logarithmically spaced;
(b) (bottom) Velocity arrows are plotted with length proportional to
gas velocity. An arrow of length equal to the spacing between arrows has 
velocity $25\,\rm km\,s^{-1}$; minimum velocities plotted are 
$3\,\rm km\,s^{-1}$. Solid line shows current extent of gas initially
inside minihalos at $z=9$. Dashed line is I-front (50\% H-ionization contour).}
\label{fig2}
\end{figure}

\section*{I-Fronts and Minihalos at High Redshift}
 
The first sources of ionizing radiation to condense out of the dark,
neutral, opaque IGM reheated and reionized it between $z\sim30$ and $z\sim6$.
Weak, R-type ionization fronts surrounding each source swept outward thru
the IGM, overtaking smaller-mass virialized halos, called
{\it minihalos}, and photoevaporating them.
High-redshift sources of ionizing photons may have found their sky covered by
these minihalos. The photoevaporation of minihalos may therefore
have dominated consumption of ionizing photons during reionization\cite{ham01}.
In this paper, we focus on the
currently favored $\Lambda$CDM model ($\Omega_0=0.3$, $\lambda_0=0.7$,
$h=0.7$, $\Omega_bh^2=0.02$). In this model, 
the universe at $z\gtrsim6$ was already filled with
minihalos capable of trapping a piece of the global,
intergalactic I-fronts, photoevaporating
their gaseous baryons back into the IGM.
Prior to their encounter with these I-fronts, minihalos
with $T_{\rm vir}<10^4\rm K$ were neutral and optically thick to hydrogen 
ionizing radiation. 
N-body simulations in a cubic volume $100\,\rm kpc$
on a side (see Fig. 1a,b) reveal that at $z=9$
minihalos with $T_{\rm vir}<10^4\rm K$ (i.e. $M<10^{7.6}M_\odot$) were
separated on average by only $d\sim7\,\rm kpc$ while their geometric
cross section together covered $\sim16\%$ of the area along every 
$100\,\rm kpc$
of an average line of sight \cite{smi03}. If the sources of reionization
were large mass halos with $T>T_{\rm vir}$, then these
were well-enough separated that typical reionization photons were likely
to have been absorbed by intervening minihalos.

To demonstrate this in a statistically meaningful way with more dynamic
range than N-body results, we combine the Press-Schechter
(PS) prescription for deriving the average number density of halos
with the truncated isothermal sphere (TIS) 
halo model\cite{is01,sir99},
to determine which
halos are subject to photoevaporation and how common they are,
as function of their redshifts. We then compute
the fraction of the
sky $F_{\rm cover,source}$, 
as seen by a source halo of a given mass, which is covered by
opaque minihalos located within the mean volume per source halo. If halos
with $M\gtrsim10^8M_\odot$ are the reionization sources, their
minihalo covering fraction is close to unity and increases by a factor of
a few if we take account of the statistical bias by which minihalos
tend to cluster around the source halos (see Fig. 1c,d).

\section*{The Photoevaporation of Minihalos}

We have performed radiation-hydrodynamical simulations of the photoevaporation
of a cosmological minihalo overrun by a weak, R-type I-front in
the surrounding IGM, created by an external source of radiation\cite{s01}. 
Minihalos are modeled as TIS
of CDM + baryons \cite{is01,sir99}. We consider 3 different source spectra:
(1) QSO-like: $F_\nu\propto\nu^{-1.8}$ ($\nu>\nu_H$);
(2) Stellar blackbody: $T_{\rm eff}=50,000\,\rm K$; (3) ``No Metals'' Stellar
$T_{\rm eff}=100,000\,\rm K$. The I-front encounters the minihalo at
redshifts
$z_{\rm initial}=(6,7,9,11)$. The flux levels are
$F_0=N_{\rm ph,56}(\nu>\nu_H)/r_{\rm Mpc}^2=(0.1,0.5,1,2,5,10,1000)$
(where $N_{\rm ph}=N_{\rm ph,56}\times10^{56}$ ionizing photons 
$\rm cm^{-2}s^{-1}$, and $r=r_{\rm Mpc}\times 1\,\rm Mpc$). 
The halo masses
are $M_{\rm halo}=(10^4,10^5,10^6,10^7,2\times10^7,4\times10^7)M_\odot$
(where $4\times10^7M_\odot$ corresponds to $T_{\rm vir}=10^4K$ at $z=9$).
The minihalo shields itself against ionizing photons, traps the R-type
I-front which enters the halo, causing it to decelerate inside the halo
to close to the sound speed of the ionized gas and transforms itself into
a D-type front, preceded by a shock. The side facing the source expels 
a supersonic wind backward toward the source, which shocks the IGM outside
the minihalo. The wind grows more isotropic with time as the remaining
neutral halo material is photoevaporated. Since the gas itself was
initially bound to a dark halo with $\sigma_V<10\,\rm km\,s^{-1}$,
photoevaporation proceeds unimpeded by gravity. Figure~2 shows the structure
of the photoevaporation flow 60 Myr after the global I-front first overtakes
a $10^7M_\odot$ minihalo.

The evaporation time per halo, $t_{\rm ev}$, and the number, $\xi$, of
ionizing photons absorbed per minihalo H atom during this evaporation
time have important implications for the reionization of the universe. We
summarize in Figure~3 the dependence of these quantities on the value of the
$M_{\rm halo}$, $z_{\rm initial}$, $F_0$,
and the source spectrum, based
on our simulation results.

\begin{figure}[b!] 
\vspace{-50pt}
  \includegraphics[width=4.6in]{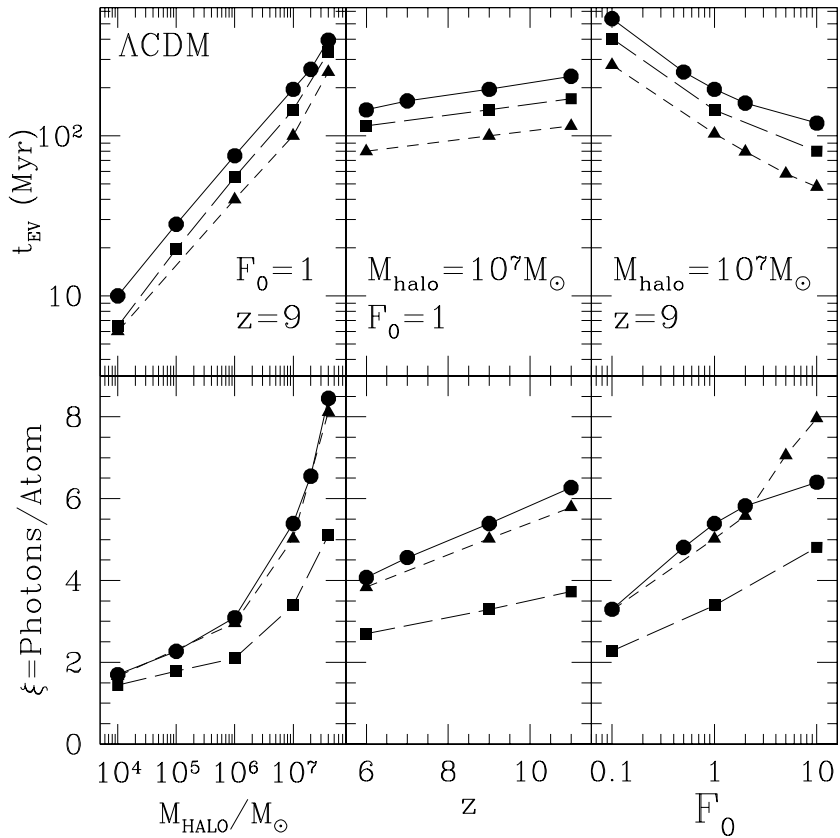}
\vspace{0pt}
\caption{Photoevaporation times $t_{\rm ev}$ for individual 
minihalos (top panels) and total number of ionizing photons absorbed
per minihalo H atom during this photoevaporation (bottom panels). 
Different panels show
the variation in these quantities with the input parameters which label
the horizontal axes: $M_{\rm halo}=$~minihalo mass, $z=$~redshift when
I-front first encounters minihalo, and 
$F_0=$~dimensionless ionizing photon flux, with
spectrum which is either a stellar blackbody, with $T_*=50,000\rm K$,
(circles) or $T_*=100,000\rm K$ (squares), or 
else QSO-like (triangles).
}
\label{fig3}
\end{figure}

%

\end{document}